\documentclass[pre,twocolumn,]{revtex4-2}

\usepackage{amsfonts}       
\usepackage{xcolor}         
\usepackage{graphicx}
\usepackage{amsmath,amssymb}

\newcommand{\be}{\begin{eqnarray}} 
\newcommand{\ee}{\end{eqnarray}}
\newcommand{\D}{\mathrm{d}}
\renewcommand{\vec}{\mathbf}

\newcommand{\bxi}{\boldsymbol\xi}
\newcommand{\blam}{\boldsymbol\Lambda}

\newcommand{\bpi}{\boldsymbol\pi}

\newcommand{\SI}{Appendix}

\renewcommand{\eqref}[1]{Eq.~({\ref{#1}})}
\newcommand{\eqqref}[2]{Eqs.~({\ref{#1}},{\ref{#2}})}

\begin{document}

\title{An effective potential for generative modelling with active matter}

\author{Adrian Baule$^1$}

\affiliation{$^1$School of Mathematical Sciences, Queen Mary University of London, London E1 4NS, United Kingdom}

\email{a.baule@qmul.ac.uk}

\begin{abstract}
Score-based diffusion models generate samples from a complex underlying data distribution by time-reversal of a diffusion process and represent the state-of-the-art in many generative AI applications. Here, I show how a generative diffusion model can be implemented based on an underlying active particle process with finite correlation time. Time reversal is achieved by imposing an effective time-dependent potential on the position coordinate, which can be readily implemented in simulations and experiments to generate new synthetic data samples driven by active fluctuations. The effective potential is valid to first order in the persistence time and leads to a force field that is fully determined by the standard score function and its derivatives up to 2nd order. Numerical experiments for artificial data distributions confirm the validity of the effective potential, which opens up new avenues to exploit fluctuations in active and living systems for generative AI purposes. 
\end{abstract}


\maketitle

Generative diffusion models are transformative developments in generative artificial intelligence (AI) for producing high-quality, diverse, and realistic images, videos, and other content. Inspired by nonequilibrium thermodynamics \cite{Sohl-Dickstein:2015aa}, they represent a powerful paradigm through which new samples from a data distribution can be generated by successively adding noise to the distribution and then reversing this process. While standard formulations of DMs are based on Gaussian perturbations of the data with varying noise scales, e.g., in denoising diffusion probabilistic models \cite{Sohl-Dickstein:2015aa,Ho:2020aa}, denoising score-matching models \cite{Vincent:2011aa,Song:2019aa}, and their continuous-time formulations \cite{Song:2021aa}, recent work has considered generalizations of the underlying diffusion process. Apart from DM approaches based on various types of non-Gaussian noises \cite{Yoon:2023aa}, recent physics-inspired work has suggested to introduce an auxiliary ``velocity" variable in an extended phase space to improve performance, e.g., using critically-damped Langevin dynamics \cite{Dockhorn:2022aa}. In \cite{Lamtyugina:2024aa}, correlated noise sources such as those of active particle processes have been explored, which map naturally onto such an extended representation. The time-reversal implementation of these approaches then introduces a vector field acting on the velocity coordinate in order to drive the generative process and has been shown to outperform standard DMs in tasks such as image synthesis and mode selection \cite{Dockhorn:2022aa,Lamtyugina:2024aa}. In the following, I show that an equivalent formalism can be derived that implements the generative process by imposing an effective potential directly on the position coordinate of an active particle while its velocity process is unchanged. The advantages of this approach are twofold. On the one hand, as I show below, the effective potential depends only on the standard score function and can thus exploit existing computational toolboxes with minimal extra effort. On the other hand, this opens up the exciting possibility to implement an {\em experimental} sampling of the generative process by imposing the potential on active particles in suspension, e.g., with optical tweezers. Such an approach could directly exploit active fluctuations for generative AI tasks suggesting novel routes for nature-inspired computing.

Generative diffusion models sample from a data distribution $p_{\rm data}$ by modelling an inverse time process. In the standard formulation using stochastic differential equations (SDEs), the forward (or noising) process is specified as \cite{Song:2021aa}
\begin{align}
\label{forward}
\dot{\vec{Y}}(t)=\vec{f}(\vec{Y},t)+g(t)\bxi(t)
\end{align}
where $\vec{f}$ is the drift and the initial position is sampled from the data distribution $\vec{Y}(0)\sim p_{\rm data}(\vec{x})$, which is assumed continuous with sample space $\vec{x}\in \mathbb{R}^d$. The noise $\boldsymbol\xi(t)$ is Gaussian white noise with covariance $<\xi_i(t)\xi_j(t')>=2D\delta_{ij}\delta(t-t')$ and $g(t)$ thus specifies the variation of noise scales, which are usually taken as monotonically increasing with $g(0)=0$ \cite{Song:2021aa}. The forward process is distributed as $\vec{Y}(t)\sim p(\vec{x},t)$, with the probability density function (PDF) determined by $p(\vec{x},t)=\int\D \vec{x}'p(\vec{x},t|\vec{x}')p_{\rm data}(\vec{x}')$ where $p(\vec{x},t|\vec{x}')$ denotes the conditional PDF of the forward process \eqref{forward} for the initial condition $\vec{Y}(0) = \vec{x}'$.

The generative process then starts from  $\vec{X}(0)\sim p(\vec{x},T)$ and generates samples of $p_{\rm data}$ using time-reversal of $\vec{Y}(t)$. The time-reversed process is given by \cite{Anderson:1982aa}
\begin{align}
\label{reverse}
\dot{\vec{X}}(t)=&-\vec{f}(\vec{X},T-t)+2D\,g^2(T-t){S}(\vec{X},t)\nonumber\\
&+g(T-t)\bxi(t)
\end{align}
where the central ingredient is the {\it score function} $\vec{S}(\vec{x},t)=\nabla\log\,p(\vec{x},T-t)$, which directs the $\vec{X}$-process towards $p_{\rm data}$, and contains the PDF $p$ of the forward process. Describing the backwards evolution in terms of a PDF $\vec{X}(t)\sim \tilde{p}(\vec{x},t)$ and starting from $\tilde{p}(\vec{x},0)=p(\vec{x},T)$, we have $\tilde{p}(\vec{x},t)=p(\vec{x},T-t)$, i.e., the score function ensures that $\vec{X}$ at any time $t$ is equal in distribution to $\vec{Y}$ at time $T-t$. The sampling of the starting point $\vec{X}(0)\sim p(\vec{x},T)$ can be simplified by suitable choices of $\vec{f},g$ such that $p(\vec{x},T)$ is well approximated, e.g., by a simple Gaussian for sufficiently large $T$ \cite{Song:2021aa}.

In the following, I consider as forward process an active particle dynamics such as an active Ornstein-Uhlenbeck particle (AOUP) with relaxation time $\tau$, where \cite{Bonilla:2019aa,Martin:2021aa}
\begin{align}
\label{aoup1}
\dot{\vec{Y}}(t)&=\vec{V}(t)\\
\tau\dot{\vec{V}}(t)&=-\vec{V}(t)+\bxi(t)\label{aoup2}
\end{align}
Compared with \eqref{forward}, the active forward process evolves in $\vec{Y}$-$\vec{V}$-space. Initial conditions are $\vec{Y}(0)\sim p_{\rm data}(\vec{x})$ and I choose $\vec{V}(0)\sim\mathcal{N}(0,D/\tau)$ such that $\vec{V}(t)$ is equivalent to a stochastic force with colored noise intensity $\left<V_i(t)V_j(t')\right>=\delta_{ij}\frac{D}{\tau}e^{-|t-t'|/\tau}$. Time reversal of \eqqref{aoup1}{aoup2} is given by \cite{Bonilla:2019aa,Lamtyugina:2024aa}
\begin{align}
\label{aouprev1}
\dot{\vec{X}}(t)&=\tilde{\vec{V}}(t)\\
\tau\dot{\tilde{\vec{V}}}(t)&=-\tilde{\vec{V}}(t)-\frac{2D}{\tau}\nabla_{\vec{v}}\log p(\vec{X},\tilde{\vec{V}},T-t)+\bxi(t)\label{aouprev2}
\end{align}
where $p(\vec{x},\vec{v},t)$ is the joint PDF of the $\vec{Y}$-$\vec{V}$ forward process \eqqref{aoup1}{aoup2}. While \eqqref{aouprev1}{aouprev2} can be used as generative process after estimation of the extended score function $\nabla_{\vec{v}}\log p(\vec{x},\vec{v},t)$, this representation is not useful in some applications, since the score function acts on the $\tilde{\vec{V}}$-variable, which is not easily controlled, e.g., in experiments. How can we instead drive the generative process by a force in the $\vec{X}$-variable only? Crucially, while \eqqref{aouprev1}{aouprev2} are the correct time-reversal equations, they are not the only possible representations of the generative process. Indeed, there are many different processes that could achieve this, since the generative process is merely constrained by the requirement to generate the correct marginal PDF $\tilde{p}(\vec{x},t)=p(\vec{x},T-t)$ as it evolves. I am interested in the following representation:
\begin{align}
\label{goal}
\dot{\vec{X}}(t)= D\blam(\vec{X},t)+\vec{V}(t),\qquad \blam(\vec{x},t)=-\nabla\mathcal{V}(\vec{x},t)
\end{align}
where $\mathcal{V}$ is an effective potential and $\vec{V}(t)$ the forward-in-time velocity of an AOUP \eqref{aoup2}. \eqref{goal} would allow sampling of the generative process by imposing a time-dependent potential $\mathcal{V}$ onto an active particle.

In order to derive $\mathcal{V}$, I first note that the linear form of \eqqref{aoup1}{aoup2} leads directly to a representation of the marginal $\vec{Y}$-process: since the joint $\vec{Y}$-$\vec{V}$-process is Gaussian, so is the $\vec{Y}$-process and thus fully specified by its first two moments. Direct integration with \eqqref{aoup1}{aoup2} yields $\mathbb{E}[\vec{Y}(t)]=\vec{Y}(0)$ and
\begin{align}
\label{variance}
\mathbb{E}[(\vec{Y}(t)-\vec{Y}(0))^2]&=2D\int_0^t \D t'\gamma(t')\nonumber\\
&=2D\left(t+\tau\left(e^{-t/\tau}-1\right)\right)
\end{align}
with $\gamma(t)=(1-e^{-t/\tau})$. As a consequence, the $\vec{Y}$-process can be described by \eqref{forward} with $\vec{f}=0$ and $g(t)=\sqrt{\gamma(t)}$. This representation highlights that \eqqref{aoup1}{aoup2} naturally implement a denoising score matching schedule as in \cite{Song:2019aa} due to the monotonically increasing variance \eqref{variance}. In addition, we see that an alternative form of the generative process is given by \eqref{reverse}, which becomes
\begin{align}
\label{reverse2}
\dot{\vec{X}}(t)=2D\,\gamma(T-t)\vec{S}(\vec{X},t)+\sqrt{\gamma(T-t)}\bxi(t)
\end{align}
Here, the initial condition is sampled as $\vec{X}(0)\sim \tilde{p}(\vec{x},0)=p(\vec{x},T)\approx \mathcal{N}(0,2D\,T)$. The latter approximation is valid for sufficiently large $T$  due to the Gaussian form of $p(\vec{x},t|\vec{x}')$ \cite{Song:2021aa}.

While \eqref{reverse2} is an exact representation of the generative process, its noise term is not equivalent to an active particle process that runs forward in time. Since the sought after \eqref{goal} is non-Markovian, it is clear that mapping \eqref{reverse2} to \eqref{goal} can not be exact. The strategy is thus to map \eqref{reverse2} instead to a Markovian approximation of \eqref{goal}. Well-known approximation schemes are the unified colored noise (UCN) and the Fox approximations, which have been widely-used to investigate effective interactions in active particle systems \cite{Maggi:2015aa,Farage:2015aa,Rein:2016aa,Martin:2021aa}. In the remainder, I focus on the Fox approximation, while results for the UCN approximation are derived in the \SI. It is useful to introduce the potential $\mathcal{V}_0(\vec{x},t)=-\log\,\tilde{p}(\vec{x},t)$ such that the score function in \eqref{reverse2} is given by $\vec{S}(\vec{x},t)=-\nabla \mathcal{V}_0(\vec{x},t)$. For simplicity, I will also assume that $\vec{Y}$ and $\vec{X}$ are made dimensionless by rescaling with a suitable scale of $p_{\rm data}$. The only remaining characteristic scales are then the time scales $D^{-1}$ and $\tau$. 

Derivations of the multivariate Fox approximation have been presented, e.g., in \cite{Farage:2015aa,Rein:2016aa,Martin:2021aa}, but for $\mathcal{V}$ given as a time-independent potential. I show in the \SI$ $ that this result remains unchanged for \eqref{goal} assuming similar approximations. The Markovian Fokker-Planck equation (FPE) associated with \eqref{goal} in the Fox approximation is thus given by
\be
\label{fpfox}
\frac{\partial}{\partial t}\tilde{p}=-D\partial_i\Lambda_i\tilde{p}+D\partial_i\partial_jG_{ij}\tilde{p}
\ee
where $\mathbf{G}=\boldsymbol\Gamma^{-1}$ and $\boldsymbol\Gamma$ has entries $\Gamma_{ij}=\delta_{ij}-D\tau\partial_j\Lambda_i$.

\begin{figure}
\includegraphics[width=\columnwidth]{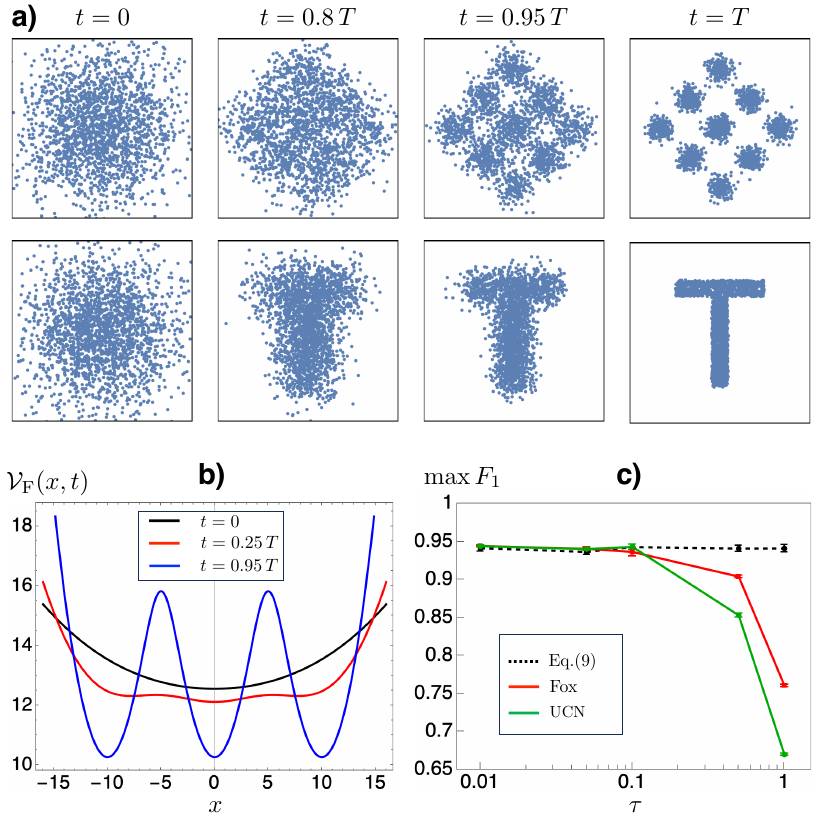}
\caption{\label{Fig:data}a) Time evolution of $2\cdot 10^3$ independent samples generated by the process \eqref{goal} for two target distributions $p_{\rm data}$. b) Plots of $\mathcal{V}_{\rm F}$ of \eqref{effpotfox} for $p_{\rm data}$ given by the Gaussian mixture for $y=0$ and different $t$. c) Quantitative evaluation of the accuracy of the generated distribution compared with the Gaussian mixture target distribution. Values are averaged over 5 repetitions with error bars given as the standard error. Parameters: $D=1$, $T=30$, time step $dt=\min(0.01,\tau/20)$.}
\end{figure}

In the next step, I manipulate the FPE for \eqref{reverse2} such that it assumes the form of \eqref{fpfox}. Introducing $\tilde{\gamma}_t=\gamma(T-t)$, the FPE for \eqref{reverse2} is
\begin{align}
\label{fprev}
\frac{\partial}{\partial t}\tilde{p}=-2D\tilde{\gamma}_t\partial_i{S}_i\tilde{p}+D\tilde{\gamma}_t\partial_i\partial_i\tilde{p}=-D\tilde{\gamma}_t\partial_i{S}_i\tilde{p}
\end{align}
since ${S}_i\tilde{p}=\partial_i\tilde{p}$. For any $G_{ij}$ we likewise have the identity
\be
\label{identity}
0&=&-\partial_i\tilde{p}\left(\partial_jG_{ij}+G_{ij}{S}_j\right)+\partial_i\partial_jG_{ij}\tilde{p}
\ee
Adding \eqref{identity} in \eqref{fprev} yields the FPE for the generative process in the form of \eqref{fpfox} and one can identify
\be
\label{lambda}
\Lambda_i&=&\partial_jG_{ij}+G_{ij}{S}_j+\tilde{\gamma}_t{S}_i.
\ee
Here, the $G_{ij}$ entries depend themselves on $\Lambda_i$ such that \eqref{lambda} is a self-consistent equation determining $\Lambda_i$. A solution can be found perturbatively. Introducing the dimensionless parameter $\alpha=D\tau$, we obtain to first order in $\alpha$: $G_{ij}\approx \delta_{ij}+\alpha\, \partial_j\Lambda_i$. Thus \eqref{lambda} becomes
\be
\label{lambda2}
\Lambda_i&=&(1+\tilde{\gamma}_t){S}_i+\alpha\left(\partial_j\partial_j\Lambda_i+{S}_j\partial_j\Lambda_i\right)
\ee
Substituting the ansatz $\Lambda_i=\Lambda_i^{(0)}+\alpha\,\Lambda_i^{(1)}+\mathcal{O}(\alpha^2)$ and neglecting terms $\mathcal{O}(\alpha^2)$ yields the solutions $\Lambda_i^{(0)}=(1+\tilde{\gamma}_t){S}_i$ and $\Lambda_i^{(1)}=(1+\tilde{\gamma}_t)(\partial_j\partial_j{S}_i+{S}_j\partial_j{S}_i)$. Since the score function has the form ${S}_i= -\partial_i\mathcal{V}_0$, we have ${S}_j\partial_j{S}_i={S}_j\partial_i{S}_j=\frac{1}{2}\partial_i{S}_j{S}_j$. At first order in $\alpha=D\tau$, the force field of \eqref{goal} in the Fox approximation is 
\be
\label{lamfox}
\blam_{\rm F}&=&(1+\tilde{\gamma}_t)\left(\vphantom{\frac{1}{2}}\vec{S}+D\tau\nabla^2\vec{S}+\frac{D\tau}{2}\nabla\left(\vec{S}\right)^2\right)
\ee
and the associated effective potential is
\be
\label{effpotfox}
\mathcal{V}_{\rm F}&=&(1+\tilde{\gamma}_t)\left(\vphantom{\frac{1}{2}}\mathcal{V}_0+D\tau\nabla^2\mathcal{V}_0-\frac{D\tau}{2}\left(\nabla\mathcal{V}_0\right)^2\right)
\ee

The validity of the effective potential is tested in numerical experiments for two types of target distributions $p_{\rm data}$ in $d=2$: (i) a mixture of 9 equal Gaussians with variance $1$ arranged in a diamond shape; (ii) a uniform distribution arranged as the letter `T'. For both cases $\mathcal{V}_{\rm F}$ can be straightforwardly evaluated in closed analytical form, see \SI. Fig.~\ref{Fig:data}a) shows plots of independent samples generated with \eqref{goal}, where the initial condition is distributed as $\vec{X}(0)\sim \mathcal{N}(0,2D\,T)$. The sequences illustrate how samples are initially drawn from a simple Gaussian and converge to the target distribution as $t\to T$. This convergence is achieved by the corresponding time variation in $\mathcal{V}_{\rm F}$, which starts from a simple quadratic form (due to $\mathcal{V}_0(\vec{x},0)\propto \vec{x}^2$) and converges to a multi-modal form according to $p_{\rm data}$ (see Fig.~\ref{Fig:data}b). The accuracy of the generated distribution compared with the target distribution is further evaluated with the $F_1$ metric \cite{Sajjadi:2018aa}, see \SI. Fig.~\ref{Fig:data}c shows the maximal $F_1$ values for samples generated with \eqref{goal} for the Fox (\eqref{effpotfox}) and UCN approximations, Eq.~[S18] derived in the \SI. Included are results for samples generated by the exact representation \eqref{reverse2} that is valid for all $\tau$ but not driven by an AOUP. We can see that \eqref{goal} with either approximation achieves the same accuracy as \eqref{reverse2} for $\tau \le 0.1$ confirming the validity of the effective potentials to first order in $\tau$. Note that the discrepancy of $\max F_1$ from the theoretical value $1$ is due to discretization errors, see \SI.

To conclude, I have presented a generative modelling framework that is able to exploit the fluctuations of an active particle process. While the derivation presented here applies to an AOUP, other active processes modelled, e.g., by run-and-tumble particles or active Brownian particles are likewise characterized by effective colored noises in their equation of motion and could thus be captured by the same formalism. It is also straightforward to generalize \eqref{aoup1} to include additional thermal noise in the generative process. Instead of using an analytical $\vec{S}$, it is of course possible to estimate the score function from data by minimising the denoising score matching loss function \cite{Vincent:2011aa,Song:2021aa}. Indeed, one advantage of the approach presented here compared with that of \cite{Lamtyugina:2024aa} is that \eqref{goal} with $\blam_{\rm F}$ of \eqref{lamfox} only depends on the conventional score function $\vec{S}$ and its derivatives, while \eqref{aouprev2} depends on $\nabla_{\vec{v}}\log p(\vec{x},\vec{v},t)$. Since standard parametrizations of $\vec{S}$ with neural networks are typically differentiable, the computation of $\blam_{\rm F}$ can make use of efficient existing algorithms. Clearly, implementing \eqref{goal} in this way vastly increases the flexibility of the modelling approach and will be considered in future work.


\newpage

\onecolumngrid

\begin{appendix}

\section{The Fox approximation for time-dependent forces}

In order to extend the multivariate Fox approximation to time-dependent forces, I follow the steps outlined in \cite{Farage:2015aa,Rein:2016aa}. First the evolution equation for $\tilde{p}(\vec{x},t)=\left<\delta(\vec{x}-\vec{X}(t))\right>$ associated with \eqref{goal} can be formally expressed as
\be
\label{fpeformal}
\frac{\partial}{\partial t}\tilde{p}=-D\partial_i\Lambda_i\tilde{p}-\partial_i\left<V_i(t)\delta(\vec{x}-\vec{X}(t))\right>
\ee
Secondly, the Furutsu-Novikov-Donsker theorem allows us to express the expectation over the product of a Gaussian function and its functional in terms of the correlation function, i.e.,
\be
\left<V_i(t)\delta(\vec{x}-\vec{X}(t))\right>&=&\int\D s\,C(t-s)\left<\frac{\delta}{\delta V_i(s)}\delta(\vec{x}-\vec{X}(t))\right>\nonumber\\
&=&-\int\D s\,C(t-s)\partial_k\left<\frac{\delta X_k(t)}{\delta V_i(s)}\delta(\vec{x}-\vec{X}(t))\right>\label{novikov}
\ee
Thirdly the functional derivative $\frac{\delta X_k(t)}{\delta V_i(s)}$ can be evaluated by solving the first-order equation
\be
\frac{\D}{\D t}\frac{\delta X_k(t)}{\delta V_i(s)}&=&\frac{\delta \dot{X}_k(t)}{\delta V_i(s)}\nonumber\\
&=&D(\partial_l\Lambda_k)\frac{\delta X_l(t)}{\delta V_i(s)}+\delta_{ik}\delta(t-s)\label{deriv}
\ee 
using \eqref{goal}. In particular, no extra terms appear here in the presence of time-dependent $\blam$. The solution of \eqref{deriv} for $t>s$ is the matrix exponential
\be
\label{derivsol}
\frac{\delta X_k(t)}{\delta V_i(s)}=\left[e^{D\int_{s}^t\D u\,\partial\blam(u)}\right]_{ik}
\ee
where $\partial\blam(t)$ has entries $[\partial\blam(t)]_{ij}=\partial_j\Lambda_i(\vec{X}(t),t)$. Substituting \eqref{derivsol} into \eqqref{fpeformal}{novikov} yields an exact evolution equation without any approximations, which, however, is not closed. In order to close the equation, the main approximation is to assume
\be
\label{mainapprox}
e^{D\int_{s}^t\D u\,\partial\blam(u)}\approx e^{D(t-s)\partial\blam(t)}
\ee
such that the exponential can be taken out of the expectation in \eqref{novikov} after conditioning $\vec{X}(t)$ on $\vec{x}$. The remaining time integral can be evaluated as
\be
\int\D s\,C(t-s)e^{D(t-s)\partial\blam}&=&D\,\mathbf{G}\left(1-e^{-t\boldsymbol\Gamma/\tau}\right)\nonumber\\
&\approx& D\,\mathbf{G}\label{approx}
\ee
for large $t$, introducing the matrix $\boldsymbol\Gamma$ with entries
\be
\label{gamma}
\Gamma_{ij}=\delta_{ij}-D\tau\partial_j\Lambda_i(\vec{x},t)
\ee
and $\mathbf{G}=\boldsymbol\Gamma^{-1}$. With the approximation \eqref{mainapprox} we thus obtain the same result as for time-independent forces \cite{Rein:2016aa,Martin:2021aa}. Clearly, \eqref{mainapprox} is quite restrictive, since it relies on an assumption of slowly varying entries of $\partial\blam$ in both $\vec{x}$ and $t$ coordinates. On the other hand, it is challenging to replace \eqref{mainapprox} with a more accurate closure. The validity of the approximation will thus be assessed ad hoc by the performance of the resulting effective potential.

\section{Effective potential in the UCN approximation}

As for the Fox approximation, we need to extend the multivariate UCN approximation used, e.g., in \cite{Maggi:2015aa,Martin:2021aa} to time-dependent forces. Starting point is the equation of motion for the particle velocities $\bpi(t)=\dot{\vec{X}}(t)$ obtained by differentiating \eqref{goal}:
\be
\label{ucnstart}
\tau\dot{\pi}_i=-\pi_i+D(1+\tau\pi_j\partial_j)\Lambda_i+D\tau\partial_t\Lambda_i+\xi_i
\ee
where \eqqref{aoup2}{goal} have been substituted. The time-dependence of $\blam$ leads to the additional drift term $D\tau\partial_t\Lambda_i$ compared with the conventional framework. The UCN approximation proceeds by setting $\dot{\pi}=0$. Re-arranging terms in \eqref{ucnstart} leads to
\be
\Gamma_{ij}\pi_j=D\Lambda_i+D\tau\partial_t\Lambda_i+\xi_i
\ee
using $\Gamma_{ij}$ of \eqref{gamma}. Following the standard approach by multiplying with the inverse $\mathbf{G}=\boldsymbol\Gamma^{-1}$ and interpreting the multiplicative noise term in the Stratonovich convention \cite{Maggi:2015aa} yields the Markovian FPE for \eqref{goal} in the UCN approximation:
\be
\label{fpucn}
\frac{\partial}{\partial t}\tilde{p}&=&-D\partial_i\tilde{p}\,G_{ij}\left(\Lambda_j+\tau\partial_t\Lambda_j\right)+D\partial_iG_{ij}\partial_kG_{jk}\tilde{p}
\ee

We can now follow the procedure in the main text to derive the effective potential. The relevant identity is now
\be
\label{identity2}
0&=&-\partial_i\tilde{p}\,G_{ij}\left(\partial_kG_{jk}+G_{jk}{S}_k\right)+\partial_iG_{ij}\partial_kG_{jk}\tilde{p}
\ee
Adding \eqref{identity2} in \eqref{fpucn} yields the FPE for the generative process in the form of \eqref{fpucn} and one can identify
\be
\label{lambdaucn}
G_{ij}\left(\Lambda_j+\tau\partial_t\Lambda_j\right)&=&G_{ij}\partial_kG_{jk}+G_{ij}G_{jk}{S}_k+\tilde{\gamma}_t{S}_i
\ee
Multiplying with the matrix $\boldsymbol\Gamma$ yields
\be
\label{lambdaucn2}
\Lambda_i&=&-\tau\partial_t\Lambda_i+\partial_jG_{ij}+G_{ij}{S}_j+\tilde{\gamma}_t\Gamma_{ij}{S}_j
\ee
With $G_{ij}\approx \delta_{ij}+D\tau \partial_j\Lambda_i$ one obtains further to first order
\be
\Lambda_i&=&-\tau\partial_t\Lambda_i+(1+\tilde{\gamma}_t){S}_i+D\tau\partial_j\partial_j\Lambda_i+D\tau(1-\tilde{\gamma}_t){S}_j\partial_j\Lambda_i
\ee
The ansatz $\Lambda_i=\Lambda_i^{(0)}+\alpha\,\Lambda_i^{(1)}+\mathcal{O}(\alpha^2)$ leads to the solutions
\be
\Lambda_i^{(0)}&=&(1+\tilde{\gamma}_t){S}_i\\
\Lambda_i^{(1)}&=&(1+\tilde{\gamma}_t)(\partial_j\partial_j{S}_i+(1-\tilde{\gamma}_t){S}_j\partial_j{S}_i)-D^{-1}\partial_t(1+\tilde{\gamma}_t){S}_i
\ee
We thus obtain the force field in the UCN approximation to first order
\be
\label{lamucn}
\blam_{\rm U}=\blam_{\rm F}-(1+\tilde{\gamma}_t)\tilde{\gamma}_t\frac{D\tau}{2}\nabla(\vec{S})^2-\tau\partial_t(1+\tilde{\gamma}_t)\vec{S}
\ee
and the corresponding effective potential is given by
\be
\label{effpotucn}
\mathcal{V}_{\rm U}&=&\mathcal{V}_{\rm F}+(1+\tilde{\gamma}_t)\tilde{\gamma}_t\frac{D\tau}{2}(\nabla\mathcal{V}_0)^2-\tau\partial_t(1+\tilde{\gamma}_t)\mathcal{V}_0
\ee
using \eqref{effpotfox}. The UCN approximation has a more complicated form than the Fox approximation without leading to better performance in the numerical experiment considered here, see Fig.~\ref{Fig:data}c).

\section{Data distributions in the numerical experiments}

In Fig.~\ref{Fig:data} I show results for two target distributions $p_{\rm data}$ in $d=2$. As shown in \eqqref{effpotfox}{effpotucn}, the effective potentials contain the basic potential $\mathcal{V}_0$, which determines the standard score function $\vec{S}=-\nabla\mathcal{V}_0$ and is given by
\be
\mathcal{V}_0(\vec{x},t)&=&\log\,p(\vec{x},T-t)\\
p(\vec{x},t)&=&\int\D \vec{x}'p(\vec{x},t|\vec{x}')p_{\rm data}(\vec{x}')
\ee
For the standard AOUP of \eqqref{aoup1}{aoup2}, the conditional PDF of the marginal $\vec{Y}$-process $p(\vec{x},t|\vec{x}')$ is given by the Gaussian
\be
p(\vec{x},t|\vec{x}')=(2\pi\,\sigma^2(t))^{-d/2}e^{-\frac{1}{2\sigma^2(t)}(\vec{x}-\vec{x}')^2},\qquad\qquad \sigma^2(t)=2D\left(t+\tau\left(e^{-t/\tau}-1\right)\right)
\ee
This yields for the two $p_{\rm data}$ considered by integration:
\begin{enumerate}
\item[(i)] Mixture of 9 Gaussians with means $\vec{m}=\{(0,0),(5,5),(-5,-5),(5,-5),(-5,5),(0,10),(0,-10),(10,0),(-10,0)\}$ and variance 1:
\be
p(\vec{x},t)=\frac{1}{9}\sum_{i=1}^9\frac{e^{-\frac{(\vec{x}-\vec{m}_i)^2}{2(1+\sigma^2(t))}}}{2\pi(1+\sigma^2(t))}
\ee
\item[(ii)] Uniform distribution arranged in the form of the letter `T'. Defining: $h(z,t)=\frac{z}{|z|}\text{erf}\left(\frac{|z|}{\sqrt{2\sigma^2(t)}}\right)$
\be
p(\vec{x},t)=\frac{1}{4}\sum_{i=1}^2\left((h(x-a_i,t)-h(x-b_i,t))(h(y-c_i,t)-h(y-d_i,t))\right)
\ee
where $a_1=-1,b_1=1,c_1=-6,d_1=4,a_2=-5,b_2=5,c_2=4,d_2=6$.
\end{enumerate}
The force fields used to implement the generative process \eqref{goal} follow by calculating the appropriate derivatives of $\mathcal{V}_0$.

\section{The $F_{1}$ metric}

The accuracy of the generated distribution compared with the target distribution is evaluated with the help of a suitable metric. I choose the $F_1$ metric, which is based on an extension of the notions of precision and recall to continuous distributions \cite{Sajjadi:2018aa}. Precision measures how much of the generated distribution is contained in the original $p_{\rm data}$, while recall measures how much of $p_{\rm data}$ is covered by the generated distribution (see \cite{Sajjadi:2018aa} for further details). The value of $F_1$ is then defined as
\be
F_1=2\cdot\frac{\text{precision}\cdot\text{recall}}{\text{precision}+\text{recall}}
\ee
The discrepancy of $\max F_1$ from the theoretical value $1$ is due to: (i) the finite sample size; (ii) the discretization error in the evaluation of $F_1$ \cite{Sajjadi:2018aa}; (iii) the finite $T$ (i.e., the error in the initial distribution of $\vec{X}(0)$); and (iv) the finite time step $dt$ (i.e., the error in the numerical implementation of the SDEs \eqqref{goal}{reverse2}). For comparison, using the \verb|Mathematica| function \verb|RandomVariate[]| to generate samples from the Gaussian mixture yields a value of $\max F_1\approx 0.949$, which only slightly improves on the accuracy of the DM methods introduced here.

\end{appendix}

\end{document}